\pdfoutput=1
\documentclass[aps,nofootinbib,prd,superscriptaddress,tightenlines,notitlepage,t
wocolumn,showpacs,floatfix]{revtex4-1}

\usepackage[colorlinks]{hyperref}
\usepackage{tabularx}
\usepackage{graphicx}
\usepackage{amssymb,bm,tensor}
\usepackage{xcolor}
\usepackage{amsmath}
\usepackage[varg]{txfonts}
\usepackage{enumerate}
\usepackage[normalem]{ulem}
\usepackage{bm}

\newcommand{\AEI}{\affiliation{Max Planck Institute for Gravitational Physics
(Albert Einstein Institute), Am M\"uhlenberg 1, Potsdam D-14476, Germany}}

\newcommand{\Maryland}{\affiliation{Department of Physics, University of
Maryland, College Park, MD 20742, USA}}

\newcommand{\Ord}{\mathcal{O}}

\begin{document}

\title{Effective action model of dynamically scalarizing binary neutron stars}
\date{\today}
\author{Noah Sennett}\email{noah.sennett@aei.mpg.de}\AEI\Maryland
\author{Lijing Shao}\email{lijing.shao@aei.mpg.de}\AEI
\author{Jan Steinhoff}\email{jan.steinhoff@aei.mpg.de}\AEI

\begin{abstract}
Gravitational waves can be used to test general relativity (GR) in the highly
dynamical strong-field regime. Scalar-tensor theories of gravity are natural
alternatives to GR that can manifest nonperturbative phenomena in neutron stars
(NSs). One such phenomenon, known as dynamical scalarization, occurs in
coalescing binary NS systems. Ground-based gravitational-wave detectors may 
be
sensitive to this effect, and thus could potentially further constrain
scalar-tensor theories. This type of analysis requires waveform models of
dynamically scalarizing systems; in this work we devise an analytic model of
dynamical scalarization using an effective action approach. For the first time,
we compute the Newtonian-order Hamiltonian describing the dynamics of a
dynamically scalarizing binary in a self-consistent manner. Despite only
working to leading order, the model accurately predicts the frequency at which
dynamical scalarization occurs. In conjunction with Landau theory, our model
allows one to definitively establish dynamical scalarization as a second-order
phase transition. We also connect dynamical scalarization to the related
phenomena of spontaneous scalarization and induced scalarization; these
phenomena are naturally encompassed into our effective action approach. 
\end{abstract}
\pacs{04.80.Cc, 95.85.Sz, 97.60.Jd, 04.50.Kd, 04.25.Nx}

\maketitle


\section{Introduction}
\label{sec:intro}

Over a century of experiments have shown that general relativity (GR) very
accurately describes the behavior of gravity. The bulk of these tests have come
from measurements of gravitationally bound systems, either with electromagnetic
observations of our Solar System~\cite{Will:2014kxa} and binary
pulsars~\cite{Wex:2014nva, Kramer:2016kwa} or with gravitational-wave (GW)
observations of coalescing binary black holes~\cite{Abbott:2016blz,
TheLIGOScientific:2016src, TheLIGOScientific:2016pea, Abbott:2017vtc}.
Combined, these systems probe GR over a large phase space, with gravitational
fields whose relative strength and dynamism span many orders of
magnitude~\cite{Psaltis:2008bb, Baker:2014zba, Berti:2015itd, Yunes:2016jcc}.
However, one corner of parameter space that has not yet been directly tested is
the highly-dynamical, strong-field regime of gravity coupled to matter, which
would be reached in the merger of a neutron star (NS) binary system.

GWs from coalescing binary neutron stars (BNSs) are expected to be detected 
by
Advanced LIGO in the near future~\cite{Abbott:2016ymx}. Tests of GR are done
using Bayesian inference~\cite{TheLIGOScientific:2016src}, comparing the
relative probability that the measured data are consistent
with a GR waveform over a non-GR waveform to search for possible deviations
from GR. Waveforms in alternative theories of gravity can be written
schematically in the Fourier domain as
\begin{align}
  h(\bm\theta;f) &= h_\text{GR}(\bm\theta;f)
  \left[1+\delta \mathcal{A}(\bm\theta;f)\right]
  e^{i \delta \psi(\bm{\theta};f)} ,
  \label{eq:ppE}
\end{align}
where $f$ is the observed  GW frequency, $\bm\theta$ represents the intrinsic
(e.g., component masses, spins, etc.) and extrinsic (e.g., distance, sky
position, etc.) parameters of the binary. We have used $h_\text{GR}(\bm\theta;f)$ 
to represent the expected
waveform in GR while $\delta \mathcal{A}$ and $\delta \psi$ are the deviations
in the amplitude and phase, respectively, from GR~\cite{Arun:2006hn,
Yunes:2009ke, Li:2011cg, Agathos:2013upa}.  One makes an ansatz for $\delta
A$ as parameterized by a set of coefficients $\left\{\alpha_i\right\}$ and for
$\delta \psi$ as parameterized by another set of coefficients
$\left\{\beta_j\right\}$.  A common choice---the so-called restricted waveforms---is 
for $\delta A$ to be identically
zero while, for frequencies
corresponding to the inspiral, $\delta \psi$ is expanded in powers of the
frequency $f$ and its logarithm $\log f$~\cite{Yunes:2009ke, Li:2011cg,
Agathos:2013upa}.  For this choice, the parameters $\left\{\beta_j\right\}$ are
simply the coefficients of the power series in $f$ and $\log f$---they
measure the deviations from GR that appear at each order in a post-Newtonian
(PN) expansion of the phase.  Because this approach makes no reference to a
particular alternative theory of gravity, constraining the parameters
$\left\{\beta_j\right\}$ can simultaneously
constrain many alternative theories using appropriate mappings.

However, this theory-agnostic approach does not capture all possible deviations
from GR because it relies on the assumption that $\delta \psi(\bm\theta;f)$
admits a series expansion in $f$ and $\log f$ during the early inspiral. In this 
paper, we study a
particular class of scalar-tensor theories of gravity in which BNSs can undergo a 
phase transition known as \textit{dynamical scalarization}~\cite{Barausse:2012da};
 the GW signals from such systems cannot be expanded in a simple 
power series. Through this phenomenon, BNSs abruptly transition from a 
configuration that closely
resembles a BNS in GR to a drastically different state. Previous efforts to
model dynamically scalarizing systems have relied on phenomonological 
waveform
models or analytic approximations of the equations of
motion~\cite{Sampson:2013jpa, Palenzuela:2013hsa, Sampson:2014qqa,
Sennett:2016rwa}.  We continue these efforts in this paper by reformulating the
PN dynamics of BNSs with dynamical scalar charges in a manner analogous to 
the
treatment of dynamical tides in GR~\cite{Vines:2010ca, Steinhoff:2016rfi}.
Using this approach, we explicitly construct a two-body Hamiltonian that
incorporates dynamical scalarization; in contrast, in
Refs.~\cite{Palenzuela:2013hsa, Sennett:2016rwa}, only the PN equations of
motion were calculated. Our results comprise an important step towards
fully-consistent waveform models of dynamical scalarization and offer a clear
interpretation of the phenomenon as a phase transition.

The paper is organized as follows. In Sec.~\ref{sec:STbasics}, we provide an
overview to scalar-tensor theories and certain nonperturbative phenomena for
NSs. In Sec.~\ref{sec:EFT} we construct an effective action to model the
dynamical scalarization of BNSs. In Sec.~\ref{sec:results} we compare results
obtained from our model to previous analytic approaches and
numerical quasi-equilibrium (QE) configuration calculations. In 
Sec.~\ref{sec:Landau}, we use
our model to solidify the interpretation of dynamical scalarization as a phase
transition and then discuss possible extensions to the model. Finally, we
present some concluding remarks in Sec.~\ref{sec:Conclusions}.

Throughout the paper we use the conventions of Misner, Thorne, and 
Wheeler~\cite{Misner:1974qy} for the metric signature and Riemann tensor. 
We work in units in which the speed of light
and the bare gravitational constant in the Einstein frame are unity.

\section{Nonperturbative phenomena in scalar-tensor gravity}
\label{sec:STbasics}

Scalar-tensor theories of gravity are amongst the most natural and
well-motivated alternatives to GR~\cite{Will:2014kxa, Berti:2015itd}. We
consider the class of theories detailed in Ref.~\cite{Damour:1992we}, in which
a massless scalar field couples nonminimally to the metric. These theories are
described in the Jordan frame by the action,
\begin{align}
S&=\int d^4 x \frac{\sqrt{-\tilde g}}{16 \pi \tilde G}\left[\phi \tilde
R-\frac{\omega(\phi)}{\phi} \tilde g^{\mu \nu} \nabla_\mu \phi \nabla_\nu
\phi\right]+S_m[\tilde g_{\mu \nu},\chi],
\label{eq:Action}
\end{align}
where $\chi$ represents all of the matter degrees of freedom in the theory and
$\tilde G$ is the bare gravitational coupling constant in the Jordan frame.
Alternatively, the action can be written in the Einstein frame by performing a
conformal transformation, $g_{\mu \nu}\equiv \phi \tilde g_{\mu \nu}$, as
\begin{align}
S=&\int d^4 x \frac{\sqrt{-g}}{16 \pi }\left[R-2 g^{\mu \nu} \nabla_\mu \varphi
\nabla_\nu \varphi\right]+S_m \left[A^2(\varphi) g_{\mu
\nu}, \chi\right],\label{eq:EinsteinS}
\end{align}
where we have introduced the scalar field,
\begin{align} \label{eq:varphi}
\varphi\equiv \int d\phi\frac{\sqrt{3+2\omega(\phi)}}{2\phi} ,
\end{align}
and defined
\begin{align}
A(\varphi)\equiv \exp\left(-\int
\frac{d\varphi}{\sqrt{3+2\omega(\varphi)}}\right).
\end{align}
Varying the Einstein-frame action yields the field equations
\begin{align}
R_{\mu \nu} -\frac{1}{2} R g_{\mu \nu}&= 8 \pi T_{\mu \nu} + 2\nabla_\mu
\varphi \nabla_\nu \varphi - g_{\mu \nu} g^{\rho \sigma} \nabla_\rho
\varphi \nabla_\sigma \varphi,\label{eq:EinsteinEq}\\
\square \varphi&= 4 \pi \alpha(\varphi) T,\label{eq:KleinGordonEq}
\end{align}
where $T^{\mu \nu}\equiv 2 (-g)^{-1/2} \delta S_m/\delta g_{\mu \nu}$ is the
stress-energy tensor of matter, $T\equiv g_{\mu \nu} T^{\mu \nu}$ is its trace,
and we have introduced the coupling,
\begin{align}
\alpha(\varphi) \equiv -\frac{d \log A}{d \varphi}=(3+2\omega)^{-1/2}.
\end{align}

Much of the seminal research in scalar-tensor alternatives to GR considered the
simple choice of a constant coupling $\alpha$, corresponding to
Jordan-Fierz-Brans-Dicke theory~\cite{Jordan:1949zz, Fierz:1956zz,
Brans:1961sx}. This theory is currently well-constrained by measurements from
the Cassini probe~\cite{Bertotti:2003rm} and of binary
pulsars~\cite{Wex:2014nva, Kramer:2016kwa}; future observations by Advanced
LIGO are not expected to improve these constraints~\cite{Abernathy2011}.
Instead, in this work we consider theories whose coupling is linear in
$\varphi$,
\begin{align} \label{eq:DEFcoupling}
\alpha(\varphi)=-\beta \varphi.
\end{align}
Such theories can give rise to phenomena that are potentially detectable by
Advanced LIGO while evading the bounds set by the Cassini
probe~\cite{Damour:1993hw, Barausse:2012da}.\footnote{Cosmological
  considerations can further constrain the class of theories with the coupling
  given by Eq.~\eqref{eq:DEFcoupling}. In particular, when $\beta$ is negative,
  the theory evolves rapidly away from GR over cosmological
  timescales~\cite{Damour:1993id,Damour:1996ke,Sampson:2014qqa}; this 
evolution
  cannot be reconciled with current Solar System observations without
   fine-tuning the theory at some point in the distant
  past. One can solve this cosmological issue by generalizing the
  coupling~\eqref{eq:DEFcoupling} to a higher-order polynomial in $\varphi$,
  which causes the scalar field to evolve to a local minimum of $A(\varphi)$ rather than diverge~\cite{Anderson:2016aoi}. However, when expanded around this local minimum, the leading order term of the modified coupling $\alpha(\varphi)$ will have the opposite sign as in Eq.~\eqref{eq:DEFcoupling}, and thus such theories no longer manifest the nonperturbative scalarization phenomena that we study here~\cite{Anderson:2016aoi}. Alternatively, one can
  add a mass term for the scalar field to Eq.~\eqref{eq:EinsteinS} to evade the
  cosmological constraints on these theories~\cite{Damour:1996ke,Alby:2017dzl}.
  Neutron stars can undergo nonperturbative phenomena analogous to those we
  consider here when immersed in a constant background massive scalar
  field~\cite{Ramazanoglu:2016kul, Yazadjiev:2016pcb}. However, recent work 
has
  revealed that this background field should in fact oscillate over relatively
  short timescales in massive scalar-tensor theories~\cite{Alby:2017dzl}. It
remains to be seen whether NSs embedded in an oscillatory background scalar
field can also exhibit nonperturbative phenomena. As is commonly done in the
literature~\cite{Damour:1992we,
  Damour:1993hw, Damour:1996ke, Barausse:2012da, Shibata:2013pra,
Sampson:2014qqa, Palenzuela:2013hsa, Taniguchi:2014fqa, Sennett:2016rwa}, 
we ignore these cosmological concerns here.} In particular, for
sufficiently negative $\beta$, such theories can manifest \textit{spontaneous
scalarization}, \textit{dynamical scalarization}, and \textit{induced scalarization}.\footnote{See Refs.~\cite{Mendes:2014ufa,Palenzuela:2015ima,Mendes:2016fby} for a discussion of similar phenomena in theories with positive $\beta$.}

Before discussing these phenomena in detail, we briefly examine the structure
of NS solutions to Eqs.~\eqref{eq:EinsteinEq} and~\eqref{eq:KleinGordonEq} to
establish some useful notation. For simplicity, we consider a static matter
source. Working far from all matter, one can expand the metric about a
Minkowskian background in powers of $\epsilon\sim m_E/r\ll 1$ where $m_E$ is
the total mass (measured in the Einstein frame) using the post-Minkowskian
formalism (see Ref.~\cite{Blanchet:2013haa} and references within). To leading
order in $\epsilon$, Eq.~\eqref{eq:KleinGordonEq} reduces to the Poisson
equation on a flat background, whose solution in this region takes the generic
form,
\begin{align}
\varphi(r)=\varphi_0+\frac{Q}{r}+\Ord\left(\frac{1}{r^2}\right),\label{eq:Qdef}
\end{align}
where we have introduced a constant background field $\varphi_0$ and defined
the \textit{scalar charge} $Q$ as the scalar monopole moment of the source.

\subsection{Spontaneous scalarization}
\label{sec:spon}

Damour and Esposito-Far\`{e}se discovered that the presence of relativistic
matter in theories with negative $\beta$ can trigger an instability in the scalar
field~\cite{Damour:1993hw}. In such theories, a sufficiently compact NS can
undergo a phase transition known as \textit{spontaneous scalarization}
corresponding to the spontaneous breaking of the symmetry $\varphi\rightarrow
-\varphi$ in Eq.~\eqref{eq:EinsteinS}. Given current constraints from binary
pulsars (see below)~\cite{Antoniadis:2013pzd, Wex:2014nva, Shao:2016ezh, Kramer:2016kwa,
Shao:2017gwu}, numerical solutions to Eqs.~\eqref{eq:EinsteinEq}
and~\eqref{eq:KleinGordonEq} reveal that an isolated NS can develop a scalar
charge of order
\begin{align}
\frac{Q}{m_E}\lesssim10^{-1},\label{eq:ChargeMagnitude}
\end{align}
through spontaneous scalarization. This figure should be contrasted with a PN
prediction for this quantity,
\begin{align}
\begin{split}
\frac{Q}{m_E}&=-\beta \varphi_0 \left(1+a_1 C+a_2 C^2+\cdots \right)
\label{eq:PNcharge}\\
&\lesssim 10^{-5} \left(1+a_1 C+a_2 C^2+\cdots \right),
\end{split}
\end{align}
where the coefficients $a_i$ are of order unity and $C\equiv m_E/R$ is the
compactness of the NS~\cite{Damour:1992we}. The
drastic difference in magnitude between Eqs.~\eqref{eq:ChargeMagnitude}
and~\eqref{eq:PNcharge} indicates that the PN expansion does not predict
spontaneous scalarization. In this sense, we describe spontaneous scalarization
as \textit{nonperturbative}; loosely speaking, one must include every term in
the infinite sum in Eq.~\eqref{eq:PNcharge} to recover the phenomenon.

The best constraints on spontaneous scalarization come from timing 
measurements
of white dwarf-NS binaries~(see, e.g.,
Refs.~\cite{Damour:1996ke, Freire:2012mg, Antoniadis:2013pzd, Wex:2014nva, Shao:2016ezh,
Shao:2017gwu}). Unlike NSs, white dwarfs (WDs) are too diffuse to develop any
significant scalar charge through spontaneous scalarization.
Consequently, WD-NS binaries can emit substantial scalar dipole flux
$\mathcal{F}_\text{dip}$, which scales as
\begin{align}\label{eq:DipoleFlux}
  \mathcal{F}_\text{dip}\propto
  \left(\frac{Q_\text{NS}}{m^E_\text{NS}}-\frac{Q_\text{WD}}{m^E_\text{WD}}
\right)^2\approx
  \left(\frac{Q_\text{NS}}{m^E_\text{NS}}\right)^2,
\end{align}
where $m^E_\text{WD}$ and
$m^E_\text{NS}$ are the masses, and $Q_\text{WD}$ and $Q_\text{NS}$ are the 
scalar charges of the WD and NS, respectively. Pulsar timing
experiments are sensitive to any anomalous decrease in the orbital period of the 
binary,
and thus can constrain $\mathcal{F}_\text{dip}$ and consequently $Q_\text{NS}/
m^E_\text{NS}$; we refer readers to
Ref.~\cite{Shao:2017gwu} for the current best limits on spontaneous
scalarization from pulsar timing.

\subsection{Dynamical and induced scalarization}

More recently, a similar phenomenon, known as \textit{dynamical scalarization},
was uncovered in numerical-relativity (NR) simulations of BNSs in the same 
class of scalar-tensor
theories with negative $\beta$~\cite{Barausse:2012da, Shibata:2013pra,
Taniguchi:2014fqa}. These simulations considered binary systems composed of 
NSs
too diffuse to undergo spontaneous scalarization in isolation. As the binaries 
coalesced, it was
found that the presence of a companion allowed the NSs to scalarize abruptly,
developing scalar charges of the same order of magnitude as might occur 
through
spontaneous scalarization. A related phenomenon, known as \textit{induced
scalarization}, was also discovered~\cite{Barausse:2012da}, in which a
spontaneously scalarized star generates a scalar charge on a companion too
diffuse to scalarize in isolation. For simplicity, we primarily focus on dynamical 
scalarization in this work; however, the model we develop can be applied to 
systems that undergo induced scalarization as well.

Numerical relativity simulations show that dynamical and induced scalarization 
hasten the plunge
and merger of BNSs relative to the same systems in
GR~\cite{Barausse:2012da,Shibata:2013pra}. Two factors dictate the difference
in merger time for scalarized versus unscalarized systems: (i) an enhancement
in energy flux, and (ii) a modification to the binding energy. A scalarized BNS
system will emit energy more rapidly than an
unscalarized system; the dissipative channels available in GR (e.g., tensor
quadrupole radiation) are enhanced for bodies with scalar charge and new
channels become available (e.g., scalar dipole radiation). Modifications to the
binding energy of scalarized systems are not well understood. In
Ref.~\cite{Taniguchi:2014fqa}, the binding energy was argued to decrease (in
magnitude) in scalarized systems, prompting an earlier merger, whereas in this
paper, we argue that it should instead increase (see Sec.~\ref{sec:Comparison}
for more detail).

Advanced LIGO will be able to distinguish between the coalescence of scalarized
and unscalarized NSs provided that their scalar charges: (i) are sufficiently
large and (ii) develop early enough in the inspiral (in the case of dynamical
scalarization)~\cite{Sampson:2013jpa, Sampson:2014qqa, Shao:2017gwu}.
Observation of such scalarization would provide direct evidence for
modifications of GR in the strong-field regime; conversely, lack of evidence of
scalarization can further constrain the space of viable scalar-tensor theories.
Depending on the NS masses and equation of state (EOS) observed in 
coalescing
BNS systems, Advanced LIGO could provide constraints competitive with current
binary-pulsar limits~\cite{Shao:2017gwu}.

Searches for deviations from GR with GWs rely on accurate and faithful 
waveform
models in modified gravity. Several models of dynamical scalarization have been
proposed in the literature, but none at the level of sophistication of
waveforms in GR. The simplest of these approaches phenomenologically model 
$\delta\psi(\bm{\theta};f)$ to reproduce features expected to arise in dynamically 
scalarized systems. For example, one can model $\delta\psi(\bm{\theta};f)$ by a 
polynomial in $f$ to capture effects such as scalar dipole radiation and/or use a 
Heaviside step function to mimic the abrupt growth of scalar charge and 
hastened merger triggered by dynamical scalarization. Detectability studies
reveal that such models may be sufficient to identify dynamical scalarization
with Advanced LIGO~\cite{Sampson:2013jpa, Sampson:2014qqa}. However,
  the accuracy of phenomenological waveform models cannot be established 
\textit{a priori}. Ultimately, one must validate and/or calibrate these models using 
independent waveforms. In GR, this comparison is made with both analytic and 
NR waveforms (e.g., the IMRPhenom waveform family~\cite{Ajith:2007qp}). 
Because very few NR simulations of dynamical scalarization have been 
produced to date, one must rely solely on more sophisticated analytic models of 
this phenomenon to verify the accuracy of phenomenological models.

A more sophisticated approach towards waveform modeling, and one we shall 
pursue in
the present work, is to solve the field equations~\eqref{eq:EinsteinEq}
and~\eqref{eq:KleinGordonEq} in some perturbative fashion (see
  Sec.~\ref{sec:EFT}). The PN approximation is an example of such an approach;
  PN waveforms are useful inspiral models in their own right and also serve as
  the foundation for more refined waveform models, such as the
  effective-one-body (EOB) formalism~\cite{Buonanno:1998gg, Buonanno:2000ef}.
  Dynamical scalarization can be modeled by augmenting~\cite{Palenzuela:2013hsa}
   or resumming~\cite{Sennett:2016rwa} the
  PN dynamics in scalar-tensor gravity; such modifications are necessary
  because dynamical scalarization is a nonperturbative phenomenon in the same
  sense as spontaneous scalarization~\cite{Sennett:2016rwa}. Both of these
  analytic approaches suffer from two shortcomings. First, simulating the
  dynamics with these models requires one to solve a system of algebraic
  equations at each moment in time involving the function $m_E(\varphi)$, which
  measures the complete (nonperturbative) dependence of the NS mass on the
  scalar field in which it is immersed. Second, these approaches only model the
  dynamics at the level of the equations of motion; no rigorous formulation of
  the two-body Hamiltonian has been constructed.

In the next section, we develop a new analytic model of dynamical scalarization
that addresses these shortcomings using an effective-action approach. First,
the scalar charges $Q$ are given by roots of a system of polynomial equations;
for systems with no background scalar field $\varphi_0$, the algebraic system
reduces to a pair of cubic equations that have a closed-form solution. These
algebraic equations depend on only two new parameters per NS [as opposed to 
the
complete functions $m_E(\varphi)$]  that can be directly interpreted as the
separation at which dynamical scalarization begins and the magnitude of scalar
charge that develops. Second, the new model allows one to construct a simple
two-body Hamiltonian and thus also compute the binding energy of a binary
system. The Hamiltonian is a fundamental building block in the construction of
perturbative waveform models. For example, the binding energy, in conjunction
with the energy flux, allows one to compute the phase evolution through the
balance equation~\cite{Blanchet:2013haa}, and the Hamiltonian is the natural
starting point in constructing an EOB description of the dynamics.
Additionally, our new formulation allows for a more nuanced interpretation of 
dynamical scalarization as a phase
transition than exists in the literature and more intimately connects dynamical 
and spontaneous scalarization.

\section{Effective action with a dynamical scalar charge}
\label{sec:EFT}

We construct a model for dynamical scalarization by explicitly
  re-parameterizing the standard point-particle action for a BNS in terms
of the scalar charges of its components. This approach closely resembles
  the treatment of extended bodies in GR in terms of their multipolar
  structure; in fact, as can be seen from Eq.~\eqref{eq:Qdef}, the scalar
  charge is simply the scalar monopole moment of an extended body. The
  gravitational fields (tensor and scalar) produced by a system of compact
  bodies can be represented completely in terms of the bodies' multipoles
  through matched asymptotic expansions~\cite{Blanchet:2013haa,
  Damour:1992we}.
  In turn, these external fields affect the multipolar structure of the compact
  bodies. This response must be included into the point-particle model in some
  way. For example, a constant external tidal field $\mathcal{G}_{i_1\ldots
  i_\ell}$ will induce a multipole $\mathcal{Q}_{i_1\ldots i_\ell}$ as
  determined by the tidal deformability $\lambda_\ell$
\begin{align}
\mathcal{Q}_{i_1\ldots i_\ell}=- \lambda_\ell \mathcal{G}_{i_1\ldots i_\ell}.
\label{eq:TidalDef}
\end{align}
(See Ref.~\cite{Damour:2009vw} for more detail.) A more
sophisticated model is needed to capture dynamical tides, i.e., tidal fields
that vary on periods comparable to the relaxation timescale of the compact
body (see Refs.~\cite{Hinderer:2016eia, Steinhoff:2016rfi, Flanagan:2007ix,
Chakrabarti:2013lua} and references therein).

As can be seen from the arguments of the matter action $S_m$ in
Eq.~\eqref{eq:EinsteinS}, compact objects in scalar-tensor gravity interact
with the scalar field in conjunction with the Einstein frame metric. For
non-self-gravitating objects (i.e., test particles), this interaction is
characterized simply by $A(\varphi)$. However, the internal gravitational
interactions in self-gravitating objects can dramatically change the couplings
to the metric and scalar field; these differences represent violations of the
strong equivalence principle. As first proposed by Eardley~\cite{Eardley:1975},
the response of a body's mass monopole $m_E$ to an external scalar field can 
be
encoded into a generic function $m_E(\varphi)$. As shown
in Appendix A of Ref.~\cite{Damour:1992we}, the scalar monopole $Q$ induced 
by
an external scalar field is given by
\begin{align}
Q= - \frac{d m_E}{d \varphi}. \label{eq:QmE}
\end{align}
For bodies immersed in weak scalar fields, Eq.~\eqref{eq:QmE} reduces to a
linear relation analogous to Eq.~\eqref{eq:TidalDef}. However dynamical
scalarization occurs outside of this linear regime: the complete expression
$m_E(\varphi)$ is needed to accurately model this phenomenon.

In this section, we develop a model inspired by the treatment of non-adiabatic
tides in GR~\cite{Hinderer:2016eia, Steinhoff:2016rfi, Flanagan:2007ix,
Chakrabarti:2013lua}. We rewrite the point-particle action using $Q$ in place
of $\varphi$ and promote the scalar charge $Q$ to a dynamical degree of
freedom. We find that this action can be expressed as a simple effective action
for a dynamical scalar charge linearly coupled to an external scalar field.
The complete function $m_E(\varphi)$ is condensed into the coupling
coefficients (or ``form factors'') in the effective action. Thus, the
predictions of the model are parameterized by a small set of coefficients and
are easy to study without reference to any particular BNS system; in
contrast previous analytic models~\cite{Palenzuela:2013hsa,Sennett:2016rwa}
required the full form of $m(\varphi)$ to be predictive.

In Sec.~\ref{sec:PointParticle}, we develop the
framework for our new effective point-particle action for a single
NS and discuss possible extensions for future work. Using this approach, we 
compute the dynamics for a binary system of two point particles in 
Sec.~\ref{sec:Dynamics}.

\subsection{The effective point-particle action}\label{sec:PointParticle}

We begin with the standard model of the orbital dynamics of compact objects in
scalar-tensor gravity.  If the orbital separation is much larger than
the size of the bodies, one can represent each star as a point particle governed 
by an action
of the form~\cite{Damour:1992we, Damour:1995kt, Damour:1998jk}, 
\begin{equation}\label{SmE}
S_m = - \int d\sigma \sqrt{- u^\mu u_\mu} m_E(\varphi) ,
\end{equation}
where $z^\mu(\sigma)$ is the object's worldline parametrized by a generic
parameter $\sigma$, ${u^\mu \equiv d z^\mu / d \sigma}$ is its four-velocity,
and $m_E(\varphi)$ is its Einstein-frame mass as a function of the scalar field
along the worldline $\varphi(z^\mu)$. Inserting the source~\eqref{SmE} into
 Eq.~\eqref{eq:KleinGordonEq}, one finds that the compact object generates
  a scalar field given by
\begin{align}
\Box \varphi =& 4\pi \int d\sigma \frac{\sqrt{- u^\nu u_\nu}}{\sqrt{-g}} \frac{d
m_E}{d \varphi} \delta^{(4)}(x^\mu - z^\mu),
\end{align}
where the derivative of the mass is evaluated at $\varphi(z^\mu)$.
  Similarly, the influence of the object on the metric can be found by
  inserting Eq.~\eqref{SmE} into Eq.~\eqref{eq:EinsteinEq}.

Next, we convert the action~\eqref{SmE} from a function of the external
field $\varphi$ imposed on the body to one of the scalar charge $Q$. These two
quantities offer complementary descriptions of the local geometry of the
compact body; one can convert between the two using Eq.~\eqref{eq:QmE}. 
To rewrite the action as a function of $Q$, we adopt a method first introduced in 
Ref.~\cite{Damour:1996ke};
we define a new potential $m(Q)$ given by the Legendre transformation of
the mass $m_E(\varphi)$,
\begin{equation}\label{Legendre}
  m(Q) \equiv m_E(\varphi) + Q \varphi .
\end{equation}
Inserting this definition into  Eq.~\eqref{SmE}, the action reads
\begin{equation}\label{Sm}
S_m = - \int d\sigma \sqrt{- u^\mu u_\mu} \left[ m(Q) - Q \varphi \right] .
\end{equation}
Now we promote $Q$ to an independent degree of freedom in the model; 
variation
of the action with respect to this variable gives an additional equation of
motion in the dynamics.

The notation in Eq.~\eqref{Legendre} is intentionally suggestive; as we
will show in Sec.~\ref{sec:Dynamics}, $m(Q)$ assumes the role of the particle's
mass in the orbital dynamics rather than $m_E(\varphi)$. A natural analogy can
be drawn to thermodynamics: consider, for example, an ideal gas composed of a
fixed number of particles held at a constant temperature. The state of the
system can be described by either its pressure---an intrinsic quantity,
analogous to $\varphi$---or its volume---an extrinsic quantity, analogous to
$Q$. While the internal energy---analogous to $m_E(\varphi)$---has a natural
interpretation as the thermal energy of the gas, it is often more convenient to
use the free energy---analogous to $m(Q)$---to describe certain physical
processes. As was
discussed in Ref.~\cite{Damour:1996ke} (and will be revisited in 
Sec.~\ref{sec:Landau}), the equilibrium state for an isolated NS will
minimize the function $m(Q)$; again, this quantity plays the role of an
effective free energy of each NS in a binary system.

We expand the potential $m(Q)$ in a power series to quartic order,
\begin{equation}\label{mexpand}
m(Q) = c^{(0)} + c^{(1)} Q + \frac{c^{(2)}}{2!} Q^2 + \frac{c^{(3)}}{3!} Q^3 +
\frac{c^{(4)}}{4!} Q^4 + \Ord\left(Q^5\right) .
\end{equation}
Recall that the action~\eqref{eq:EinsteinS} equipped with the
coupling~\eqref{eq:DEFcoupling} is invariant under the symmetry
$\varphi\rightarrow - \varphi$. Thus, we expect the mass of an isolated NS
$m_E(\varphi)$ to also respect this symmetry, even in the presence of
spontaneous scalarization. From Eq.~\eqref{eq:QmE}, we see that this parity
transformation will also send $Q\rightarrow -Q$. Performing both of these
transformations leaves the right hand side of Eq.~\eqref{Legendre} unchanged,
and thus we can conclude that $m(Q)$ must be an even function of $Q$.

Some of the coefficients $c^{(n)}$ have an immediate interpretation.  The
leading $c^{(0)}$ describes the body's mass in absence of any scalar charge,
i.e., the ADM mass in GR, and so we also denote it as $c^{(0)} = m^{(0)}$.
Furthermore, a background scalar field $\varphi_0$ can be handled by
  working instead with the field,
\begin{align}
\hat\varphi\equiv \varphi-\varphi_0,
\end{align}
leading to an additional coupling $- Q \varphi_0$ in the Lagrangian. This term 
can be absorbed into $m(Q)$
 by setting $c^{(1)} = - \varphi_0$, and thus we can interpret
  $c^{(1)}$ as a cosmologically imposed background scalar field. Note that the
  addition of a nonzero scalar background weakly breaks the symmetry
  $\varphi\rightarrow -\varphi$ in the point-particle action, prompting us to
  relax the conclusion that $m(Q)$ is a strictly even function. However, all
  other odd powers of $Q$ will still vanish, i.e., $c^{(3)}=0$.
  
Given the discussion above, our model for $m(Q)$ reduces to
\begin{equation}\label{mexpand2}
m(Q) = m^{(0)} - \varphi_0 Q + \frac{c^{(2)}}{2} Q^2 + \frac{c^{(4)}}{24} Q^4 +
\Ord(Q^6) .
\end{equation}
Potentials of this form are widely used to describe systems that exhibit 
spontaneous symmetry breaking (see also Sec.~\ref{sec:Landau}); the Higgs 
mechanism is one notable example~\cite{Peskin:1995ev}. Reference~
\cite{Damour:1996ke} employed a similar potential to model isolated NSs near 
the critical point for spontaneous scalarization. In the present work, we show that 
the ansatz~\eqref{mexpand2} remains valid for NSs far from this critical point;
we describe the procedure by which we numerically compute the various
coefficients for a particular NS in Sec.~\ref{sec:match}.

One ingredient conspicuously absent from our effective action~\eqref{Sm} is the 
\textit{dynamical} response of the scalar charges to changes in the
scalar field. In truth, our model is only valid in the adiabatic limit, wherein
the external fields evolve over timescales much longer than the relaxation
time of NSs. Given the abrupt nature of dynamical scalarization, the validity
of our assumption of adiabaticity should be studied in greater detail; we
  reserve this analysis for future work. If one
rapidly changes the external scalar field, the NS's scalar charge cannot
respond instantaneously.  In general, physical systems undergo (harmonic)
oscillations around equilibrium configurations under small perturbations. Thus, 
one expects the scalar charge to behave approximately like a harmonic
oscillator driven by the external fields, characterized by an action of the
form~\eqref{Sm} with
\begin{equation}\label{Smosci}
  m(Q,\dot{Q}) = m^{(0)} - \varphi_0 Q
    + \frac{c^{(2)}}{2} \left( \frac{\dot{Q}^2}{\omega_0^2 u^\mu u_\mu} + Q^2
  \right) ,
\end{equation}
where $\dot{Q} = d Q / d \sigma$ and $\omega_0$ is the resonant frequency of
this scalar mode. This action is analogous to the dynamical tidal model in
Ref.~\cite{Steinhoff:2016rfi}: $Q$ corresponds to the dynamical quadrupole,
$\varphi$ to the tidal field, $1/c^{(2)}$ to the tidal deformability, and
$\omega_0$ to the oscillation mode frequency. In general, one should add
separate dynamical degrees of freedom for every oscillation mode of the NS.
Identifying all dynamical degrees of freedom relevant for the scale of interest
is very important in constructing an effective action (see, e.g.,
Ref.~\cite{Weinberg:2016kyd}). Note that when the dynamics of the system 
occur much more slowly than the resonant frequency, i.e. $\dot{Q} \ll \omega_0$, 
and we restore the $Q^4$ interaction, Eq.~\eqref{Smosci} reduces to the 
adiabatic model~\eqref{mexpand2} considered earlier.

Viewed from an effective field theory perspective, our effective action model
of dynamical scalarization may appear too simplistic. In
general, one should add  to the action all possible combinations of $Q$,
$u^\mu$, the scalar field $\varphi$, and the curvature (and derivatives of
these variables) allowed by the symmetries of the theory, up to terms
negligible for the desired accuracy of the model. Not all of these interactions
are independent, since some might be related by redefinitions of the other
dynamical variables; the redundant terms should be dropped. In the present
model, we consider only couplings of the scalar charge to itself, as well as a
linear coupling of the charge to the scalar field. A broader class of
interactions would allow our model to reproduce other interesting phenomena. 
For example, the induction of scalar charges on black holes from time-varying 
external fields can be modeled with an effective action~\cite{Jacobson:1999vr,
Horbatsch:2011ye}. We delay such an investigation for future work; for the
present work, the effective action model given by Eqs.~\eqref{Sm}
and~\eqref{mexpand2} is sufficient to reproduce dynamical scalarization.

\subsection{Dynamics of a binary system}\label{sec:Dynamics}

We now turn to the task of translating the action [which will contain a copy of
Eq.~\eqref{Sm} for each NS] into a Hamiltonian describing the orbital dynamics
of a BNS. Using the PN approximation, we expand the metric and scalar field in
powers of $v/c$ and solve the field equations~\eqref{eq:EinsteinEq}
and~\eqref{eq:KleinGordonEq} at each order. An efficient method for solving the
two-body dynamics is through a Fokker action\footnote{This means to insert the
  perturbative solution to the field equations into the full action.} together
  with a diagrammatic method to represent the perturbative
  expansion~\cite{Damour:1995kt}. Similarly, one can integrate out the fields
  perturbatively using techniques from quantum field
  theory~\cite{Goldberger:2004jt}, i.e., Feynman integrals and diagrams.

We consider only the leading-order (Newtonian) approximation of the orbital
dynamics in the present work. Thus, the accuracy of our model will degrade
towards the end of the inspiral. However, because Advanced LIGO is only
sensitive to dynamical scalarization that occurs in the very early
inspiral~\cite{Sampson:2013jpa,Sampson:2014qqa}, our model can still be 
applied to the systems of scientific interest;
we pursue extensions of our model
to higher PN order in future work.

The PN expansions of the metric $g_{\mu \nu}$ and the scalar field $\hat\varphi$ 
are given by
\begin{align}
g_{\mu \nu} =& \eta_{\mu \nu}+h_{\mu \nu}+\Ord\left(c^{-4}\right),\label{eq:hDef}\\
\hat\varphi =& \psi+\Ord\left(c^{-4}\right),\label{eq:psiDef}
\end{align}
where $\eta_{\mu \nu}$ is the Minkowski metric and $\hat\varphi$ vanishes at 
infinity by construction. The leading-order PN corrections enter with the following 
powers of $c$:
\begin{align}
h_{00} & \sim \Ord\left(c^{-2}\right), && h_{0i} \sim \Ord\left(c^{-3}\right),
\nonumber \\
h_{ij} & \sim \Ord\left(c^{-4}\right), && \psi \sim \Ord\left(c^{-2}\right).
\end{align}

Inserting the expansions~\eqref{eq:hDef} and~\eqref{eq:psiDef} into the
field equations~\eqref{eq:EinsteinEq} and~\eqref{eq:KleinGordonEq} with the
source~\eqref{Sm}, one finds the Newtonian-order solution to the metric and
scalar field,
\begin{subequations}
\begin{align}
h_{00}(\mathbf{x},t)&= \frac{m_A(Q_A)}{|\mathbf{x}-\mathbf{z_A}(t)|}+
\frac{m_B(Q_B)}{|\mathbf{x}-\mathbf{z_B}(t)|}+\Ord\left(c^{-4}\right),\\
\psi(\mathbf{x},t)&=\frac{Q_A}{|\mathbf{x}-\mathbf{z_A}(t)|}+\frac{Q_B}{|
\mathbf{x}-\mathbf{z_B}(t)|}+\Ord\left(c^{-4}\right),\\
h_{0i}(\mathbf{x},t)&\sim \Ord\left(c^{-3}\right),\\
h_{ij}(\mathbf{x},t)&\sim \Ord\left(c^{-4}\right),
\end{align}
\end{subequations}
where the labels $A$ and $B$ distinguish the two NSs. Henceforth, we suppress 
the explicit dependence of
each body's mass $m$ on its corresponding scalar charge for notational 
convenience.

Inserting these solutions into the action and dropping singular self-interactions, 
we find the leading-order two-body
action,
\begin{equation}\label{SPN}
  S \approx \int dt \bigg[
  - m_A - m_B
  + \frac{m_A}{2} \mathbf{v}_A^2 + \frac{m_B}{2} \mathbf{v}_B^2
  + \frac{m_A m_B}{r} + \frac{Q_A Q_B}{r} \bigg] ,
\end{equation}
where $v^i \equiv d z^i / d t$ is the Newtonian velocity and ${r \equiv
|\mathbf{z}_A - \mathbf{z}_B|}$ and we have corrected for any double counting.
Legendre transforming the Lagrangian yields the Hamiltonian,
\begin{equation} \label{Hamiltonian}
H=m_A + m_B
  + \frac{\mathbf{p}_A^2}{2 m_A} + \frac{\mathbf{p}_B^2}{2m_B}
  - \frac{m_A m_B}{r} - \frac{Q_A Q_B}{r} ,
\end{equation}
where the canonical momenta are $\mathbf{p}_{A,B}=m_{A,B} \mathbf{v}_{A,B}
$. The equation of motion for $Q_A$ reads
\begin{align}\label{QEOM2}
\begin{split}
 0 = \frac{\partial H}{\partial Q_A} = z_A \left(- \varphi_0 + c^{(2)}_A Q_A +
 \frac{c^{(4)}_A}{6}
 Q^3_A\right)
 - \frac{Q_B}{r},
\end{split}
\end{align}
with the redshift given by
\begin{align}
\begin{split}\label{eq:zDef}
  z_A = \frac{\partial H}{\partial m_A} =
  1-\frac{\mathbf{p}_A^2}{2 m_A^2}-\frac{m_B}{r} ,
\end{split}
\end{align}
and the equation of motion for $Q_B$ takes the same form but with the body
labels exchanged $A\leftrightarrow B$. The scalar charges are given by the
roots of these two cubic equations.\footnote{For consistency, we truncate
 Eq.~\eqref{QEOM2} at cubic order in the scalar charges, e.g.
 dropping the term proportional to $Q_A^3 Q_B$ that would arise from the 
product of $m_B$ and $Q_A^3$.} Closed form solutions can be found using
computer algebra for $\varphi_0 \neq 0$, but the result is rather lengthy and
not very illuminating; we do not provide them here for space considerations.

While Eq.~\eqref{QEOM2} may seem daunting,
simple analytic solutions for the scalar charge can be easily found in special,
but very relevant cases. We restrict our attention to the theories that exactly
preserve the symmetry $\varphi\rightarrow - \varphi$, i.e., we set the
background scalar field $\varphi_0=0$. Next, for simplicity, we will neglect
the $\Ord(c^{-2})$ corrections to the redshift $z_A$ in Eq.~\eqref{QEOM2}; 
including
these terms does not change the qualitative behavior of the solutions discussed
below. Finally, we specialize to the case of
equal-mass binaries and assume that the NSs
have identical properties, i.e. $m_A^{(0)} = m_B^{(0)}$ and $c^{(i)}_A = c^{(i)}_B
$.
Under these assumptions, Eq.~\eqref{QEOM2} reduces to
\begin{equation} \label{QEOM2:EqualMass}
0 = \frac{\partial H}{\partial Q} = - 2Q \left[ \frac{1}{r} - c^{(2)} -
\frac{c^{(4)}}{6} Q^2 \right] ,
\end{equation}
where we have dropped the body labels. As expected, the trivial solution $Q=0$
satisfies this equation. However, this is not necessarily the only solution; if
the trivial solution is unstable, the BNS system will transition to a state
with nonzero scalar charge. The requirement for stability,
\begin{equation}
0 \leq \frac{\partial^2 H}{\partial Q^2} = 2 c^{(2)} - \frac{2}{r} +
  c^{(4)}
Q^2 ,
\end{equation}
is violated for $Q=0$ when $1 / r > c^{(2)}$.  The stable solutions therefore
read,
\begin{equation}\label{eq:Q:solution:equalmass}
  Q = \left\{ \begin{tabular}{lll}
                $0$ & \quad\text{for} & $1 / r \leq c^{(2)}$ \\
                $\pm \sqrt{\dfrac{6}{c^{(4)}}} \sqrt{\dfrac{1}{r} - c^{(2)}}$ &
                \quad\text{for} & $1 / r \geq c^{(2)}$
                \end{tabular} \right. ,
\end{equation}
which contains a phase transition at $r_\text{DS} = 1 / c^{(2)}$.

Equation~\eqref{eq:Q:solution:equalmass} provides some intuition into the
physical interpretation of the coefficients $c^{(2)}$ and $c^{(4)}$. The
parameter $c^{(2)}$ determines the orbital scale of the phase transition to the
scalarized regime, where the scalar-parity symmetry is broken and the solution
bifurcates.  The parameter $c^{(4)}$ determines the size of the scalar charge
in this regime.  Notice that for negative $c^{(2)}$ the NS is scalarized for
all $r$. In fact, this situation corresponds to spontaneous scalarization; we
discuss the connection between spontaneous and dynamical scalarization in
greater detail in Sec.~\ref{sec:Landau}.

Finally, we compute the Newtonian-order equations of motion for each NS.
  Working from the Hamiltonian~\eqref{Hamiltonian}, the equations of motion are
  given by,
\begin{align}
  \ddot{\mathbf{z}}_A=-\frac{m_B \left(1+\alpha_A \alpha_B\right)}{r^2}
\mathbf{n} ,
\end{align}
where $\alpha_{A,B} \equiv Q_{A,B} / m_{A,B}$ and
$\mathbf{n}\equiv(\mathbf{z}_A-\mathbf{z}_B)/r$. Note that $\alpha_A$ differs 
from the quantity found in Eqs.~\eqref{eq:ChargeMagnitude}--
\eqref{eq:DipoleFlux} because it uses $m(Q)$ in place of $m_E(\varphi)$. We 
also derive Kepler's third
law for circular orbits
\begin{align}
  \Omega^2=\frac{\left(m_A+m_B\right) \left(1+\alpha_A
\alpha_B\right)}{r^3}, \label{eq:Kepler}
\end{align}
where $\Omega$ is the orbital frequency.

\section{Results}\label{sec:results}

The previous sections aimed to motivate and develop a novel analytic model of
dynamical scalarization; in this section, we test the accuracy of this approach
by comparing against previous models~\cite{Sennett:2016rwa} and numerical 
QE configuration
calculations~\cite{Taniguchi:2014fqa}. The dynamics are determined entirely by 
the
coefficients $c^{(i)}$, as can be seen by inserting Eq.~\eqref{mexpand2} and
the solution of the cubic equations~\eqref{QEOM2} for $Q_{A,B}$ into the
Hamiltonian~\eqref{Hamiltonian}. 
These coefficients characterize the behavior of each compact body {\it in
isolation}, and thus can be computed straightforwardly.

To facilitate comparison with previous work, we restrict our attention to the
binary systems considered in Refs.~\cite{Sennett:2016rwa,Taniguchi:2014fqa}. 
We
consider $(1.35+1.35)\,M_\odot$ nonspinning BNS systems with a piecewise
polytropic fit to the APR4 EOS; see Ref.~\cite{Read:2008iy} for more details on
this EOS and its polytropic fit. We examine configurations with $\beta=-4.2$
and $\beta=-4.5$, where $\beta$ characterizes the strength of the scalar
coupling~\eqref{eq:DEFcoupling}. Finally, we add the background scalar field
$\varphi_0=10^{-5} / \sqrt{-2 \beta}$, which satisfies binary-pulsar
constraints for this EOS~\cite{Shibata:2013pra}.

\subsection{Computing $c^{(i)}$}\label{sec:match}

The coefficients $c^{(i)}$ describe how the energy of an isolated NS varies
with its scalar charge $Q$. Thus, to extract these coefficients, we study the
behavior of the NS under infinitesimal changes in $Q$. In practice, we compute
sequences of NS solutions with equal baryonic mass with incremental changes 
to
the mass $m_E$, scalar charge $Q$, and asymptotic field $\varphi$. Spherically
symmetric solutions for perfect fluid stars are found by solving the
Tolman-Oppenheimer-Volkoff (TOV) equations; the extensions of these 
equations
to scalar-tensor gravity were derived in Refs.~\cite{Damour:1993hw,
Damour:1996ke}. We solve these equations using fourth order Runge-Kutta 
methods
and use standard shooting techniques to construct solutions with the same
baryonic mass. The quantities $m_E$, $Q$, $\varphi$ parameterize the 
asymptotic behavior of each numerical solution; we extract $m_E$, $Q$, $\varphi
$  using the relations detailed in Refs.~\cite{Damour:1993hw, Damour:1996ke}.
Equipped with these quantities, we then compute $m(Q)$ using
Eq.~\eqref{Legendre}.

\begin{figure}
\includegraphics[width=\columnwidth]{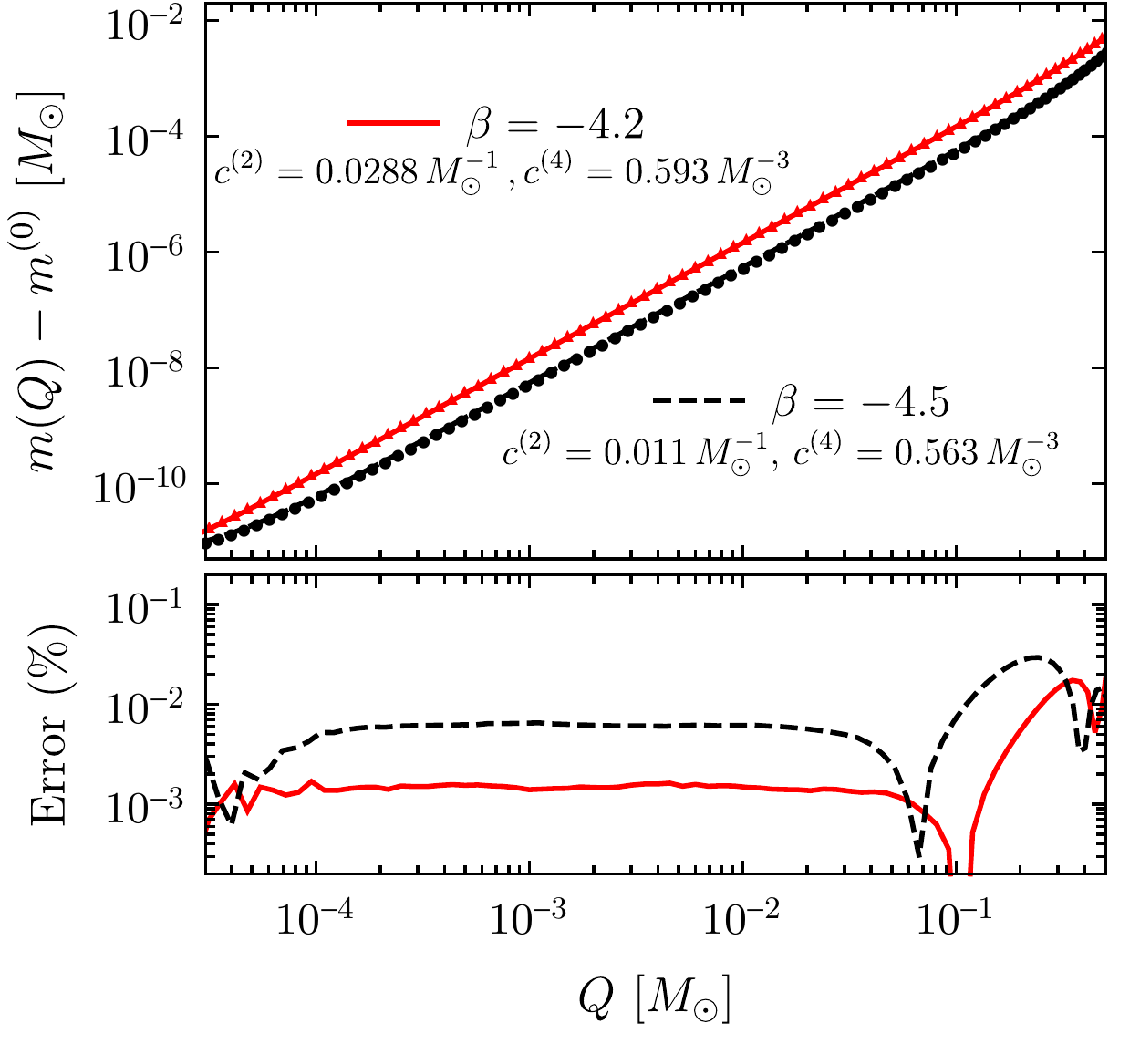}
\caption{Potential $m$ as a function of scalar charge $Q$ for a $1.35\,M_\odot$
  NS with the APR4 EOS. \textit{Top:} The numerical values and polynomial fit
  are plotted with points and lines, respectively, for $\beta=-4.2$ (red) and
  $\beta=-4.5$ (dashed black). We have subtracted the leading-order term
  $m^{(0)}=1.35\,M_\odot$ from $m$ to improve readability. \textit{Bottom:} We
  plot the fractional error in $m(Q)-m^{(0)}$ between the numerical data and
polynomial fits.}\label{fig:CoefficientFit}
\end{figure}

We compute the coefficients $c^{(i)}$ by fitting the numerically computed
$m(Q)$ with a polynomial of the form~\eqref{mexpand2}. The numerical
values and polynomial fit of $m(Q)$ are plotted with dots and solid lines,
respectively, in the top panel of Fig.~\ref{fig:CoefficientFit} for the NS
parameters discussed above. To improve readability, we have subtracted the
leading-order coefficient ${m^{(0)}=1.35M_\odot}$ from $m$. The values for
$c^{(i)}$ computed through the polynomial fit are also given in
Fig.~\ref{fig:CoefficientFit}; the $i$-th coefficient has dimension of
$\text{[mass]}^{i-1}$. The bottom panel of the figure shows the fractional
error between numerical values and polynomial fits of $m-m^{(0)}$. We see that
deviations are generally of the order $\lesssim0.01\%$, slightly worsening as the
charge increases. The range in $Q$ plotted here covers the typical range
achievable by this NS over an entire inspiral in which dynamical scalarization
occurs. As a check of our initial ansatz~\eqref{mexpand2}, we also fit the data to 
polynomials including $Q^3$, $Q^5$ and $Q^6$ terms; we find that these 
additional powers of $Q$ shift our estimates for $c^{(i)}$ by less than $\sim 0.1
\%$ and only marginally improve the overall agreement to data.

\subsection{Comparison against previous models}\label{sec:Comparison}

\begin{figure}
\includegraphics[width=\columnwidth]{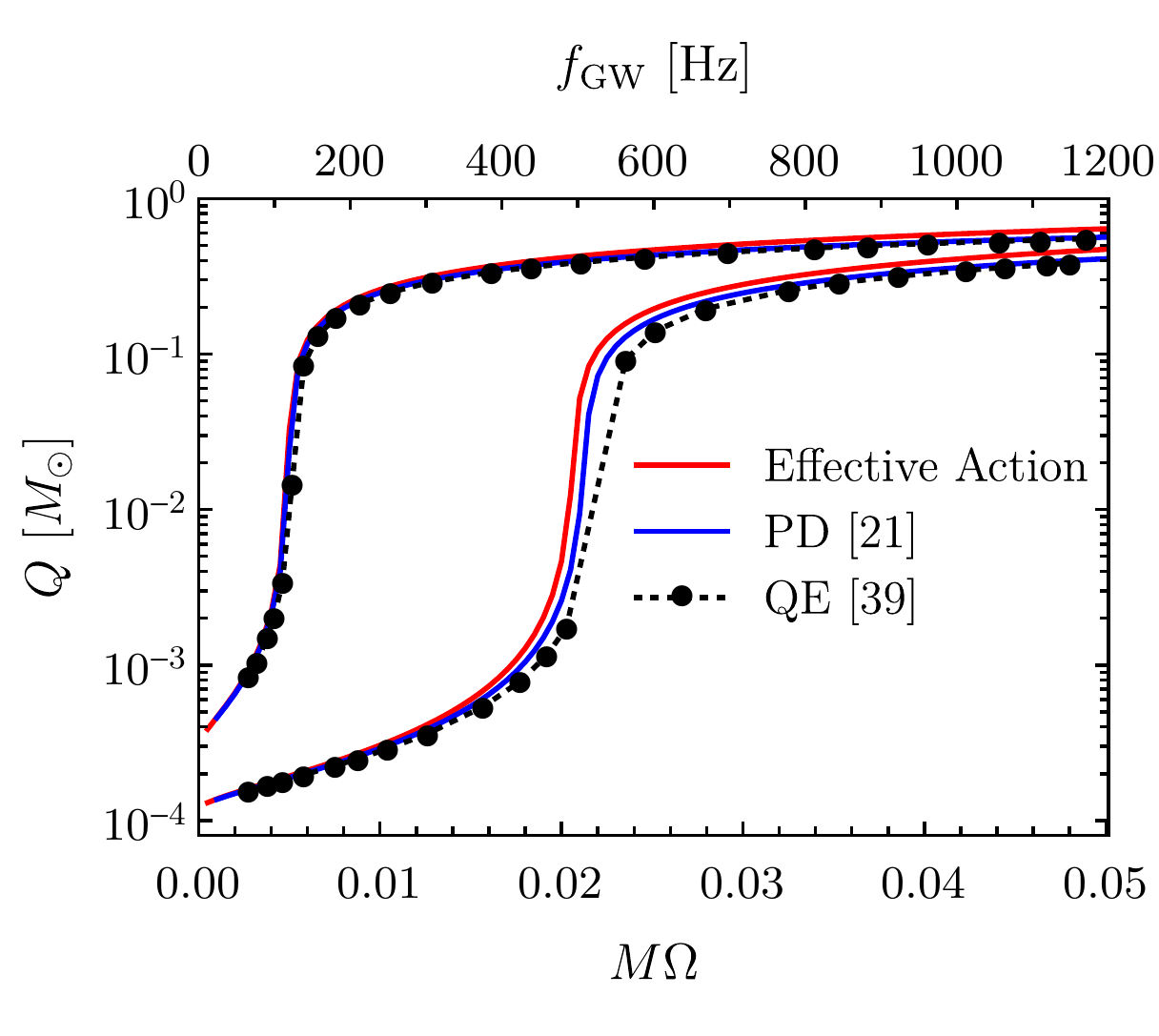}
\caption{Scalar charge of each star as a function of frequency for a
  $(1.35+1.35)\,M_\odot$ BNS with the APR4 EOS. The lower axis indicates the
  orbital frequency $\Omega$; the upper axis shows the dominant GW frequency
  $f_\text{GW}=\Omega/\pi$. The model developed here using an effective action
  is shown in red. The analytic post-Dickean (PD) model of
  Ref.~\cite{Sennett:2016rwa} is shown in blue. The numerical calculations of
  quasi-equilibrium (QE) configurations performed in
  Ref.~\cite{Taniguchi:2014fqa} are shown in black. The curves depicting earlier 
scalarization were computed with $\beta=-4.5$; the other set of curves 
correspond to $\beta=-4.2$.
}\label{fig:Qomega}
\end{figure}

As a first test of our model, we compute the scalar charge $Q$ as a function of
frequency. Because we only consider equal-mass systems, this relation can be
found by solving the cubic equation~\eqref{QEOM2} for $Q = Q_A = Q_B$ as a
function of separation $r$.  Then, by inserting this result into
Eq.~\eqref{eq:Kepler}, we determine an exact relation between $r$ and the
orbital frequency $\Omega$. Finally, we invert this relation and insert it into
the solution to Eq.~\eqref{QEOM2:EqualMass} to find an implicit expression for
$Q(\Omega)$. We plot $Q(\Omega)$ in Fig.~\ref{fig:Qomega} computed with our
model in red. The lower axis gives the dimensionless orbital frequency,
normalized by the total rest mass $M$, which we define as,
\begin{align}
M\equiv m_A^{(0)}+m_B^{(0)},
\end{align}
i.e. the sum of the component ADM masses in GR. The upper axis gives the 
dominant frequency
$f_\text{GW}=\Omega/\pi$ of the GWs produced by the binary in hertz.

We plot in blue the predictions of the
\textit{post-Dickean} (PD) model constructed in
Ref.~\cite{Sennett:2016rwa}. The PD approach resums the PN dynamics to
reproduce dynamical scalarization. To accomplish this resummation, one 
promotes the mass $m_E$ and its derivatives to functions of two scalar fields
$m_E(\varphi, \xi)$, Then, one field ($\varphi$) is integrated out of the
point-particle action~\eqref{SmE} through a standard PN expansion, while the
other ($\xi$) is treated as a new dynamical degree of freedom in the theory. In
this way, the PD approximation resembles the model presented here.  Both
methods introduce new degrees of freedom at the level of the action, and
extremizing the action with respect to these quantities yields algebraic
equations that relate the quantities to the bodies' positions and momenta.
However, in the PD approach, these equations involve the potentially
complicated function $m_E(\varphi, \xi)$ and its derivatives, whereas in the
formalism presented here, one needs only the coefficients $c^{(i)}$. In the
notation of Ref.~\cite{Sennett:2016rwa}, we define the natural analog of the
scalar charge as $Q\equiv m^{(\text{RE},\varphi)}
\alpha^{(\text{RE},\varphi)}/\sqrt{\phi_0}$ and plot this quantity in the
figure; see Eqs.~(A3) and~(A4) in Ref.~\cite{Sennett:2016rwa} for the explicit
definitions of these quantities.  The blue curve shown in Fig.~\ref{fig:Qomega}
corresponds to the next-to-leading-order dynamics in an expansion in $c^{-2}$.

Finally, we plot the results of the numerical QE
configuration calculations performed in Ref.~\cite{Taniguchi:2014fqa} with
black dots. These calculations were made under the assumption of conformal
flatness and stationarity; physically, each configuration represents a binary
on an exactly circular orbit emitting no GWs. This setup is used to approximate
a BNS during its adiabatic inspiral. The scalar mass $M_\text{S}$ of the
total system, defined in the Jordan frame, was computed in
Ref.~\cite{Taniguchi:2014fqa}. To convert this quantity to the scalar charge
of the full system, we use $Q_\text{tot}=M_\text{S}/\left(-\beta
\varphi_0\right)$; this conversion is discussed in detail in footnote 2 of 
Ref.~\cite{Taniguchi:2014fqa}. For simplicity, we assume that the component 
scalar charges are simply
half of the total scalar charge, $Q=Q_\text{tot}/2$.

As evidenced by Fig.~\ref{fig:Qomega}, we find very close agreement to previous
predictions of the evolution of the scalar charge with our effective-action
model. A key feature is the frequency $\Omega_\text{DS}$
at which dynamical scalarization occurs. As discussed above, our model predicts
the onset of dynamical scalarization when the binary separation $r=1/c^{(2)}$.
Converting the separation into an orbital frequency using
Eq.~\eqref{eq:Kepler}, we find agreement to within $\lesssim 10\%$ compared to
the values presented in Table II of Ref.~\cite{Sennett:2016rwa} for both the PD
model and the QE configuration calculations.\footnote{The agreement can be 
slightly improved by neglecting the $\Ord\left(c^{-2}\right)$ contributions to the 
redshift variables \eqref{eq:zDef} that enter into Eq.~\eqref{QEOM2}} We 
emphasize that our effective action model is in no way calibrated to fit the QE 
results; the only numerical input to the model comes from isolated NS solutions 
of the TOV equations.

\begin{figure}
\includegraphics[width=\columnwidth]{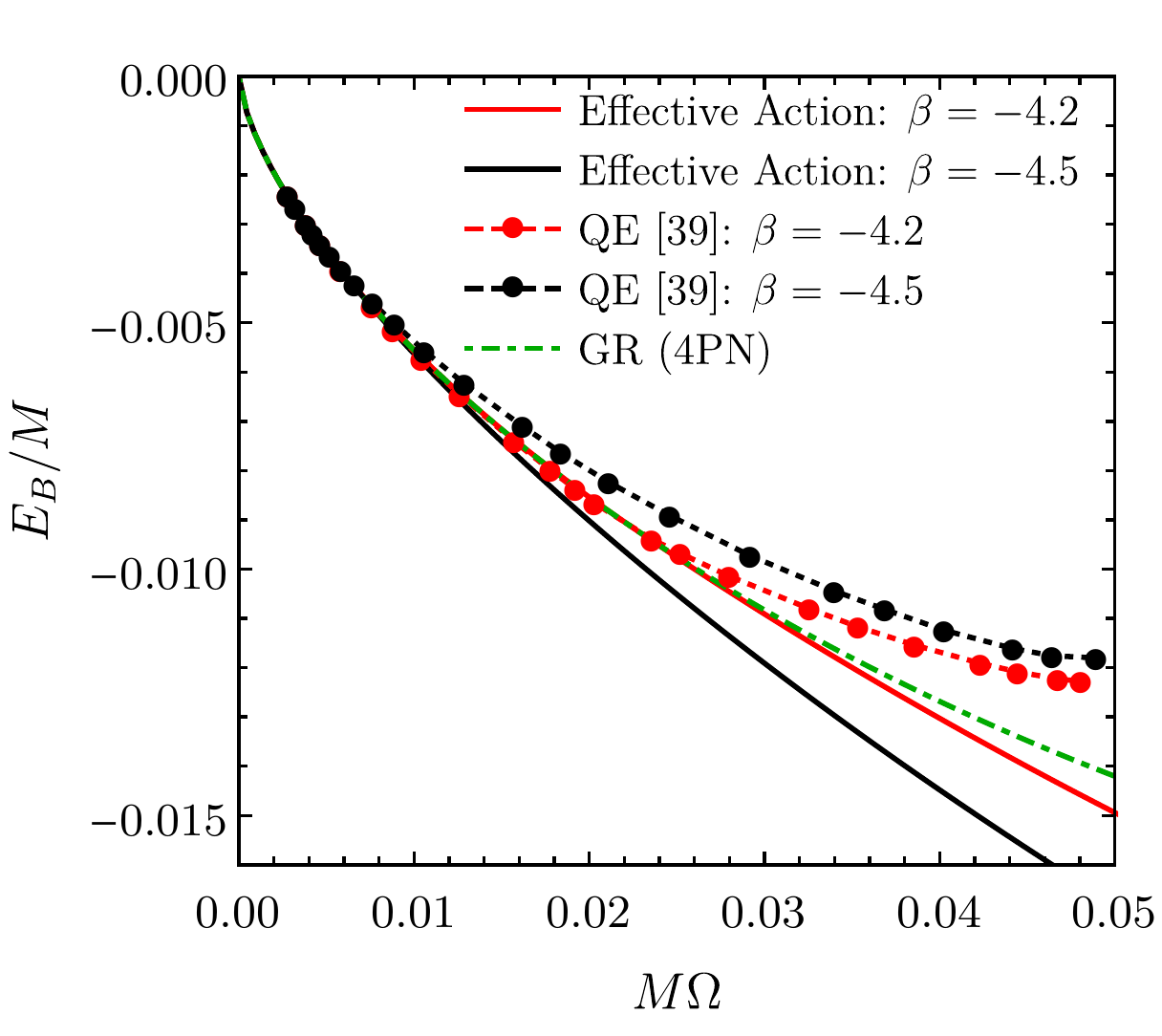}
\caption{Binding energy $E_B$ normalized by the total mass $M=2.7 M_\odot$ 
as a function of orbital frequency for the same
  BNS system as in Fig.~\ref{fig:Qomega}. The predictions of the effective action 
model
  introduced here are shown in solid lines; we add to our Newtonian-order result 
the 1PN, 2PN, 3PN, and 4PN contributions found in GR. The QE configuration
  calculations performed in Ref.~\cite{Taniguchi:2014fqa} are shown with dashed
  lines. Red (black) curves correspond to $\beta=-4.2$ ($\beta=-4.5$). For
  comparison, we have also plotted in green the 4PN prediction for  the
  point-particle binding energy in GR.
}\label{fig:Eomega}
\end{figure}

Having computed $Q(\Omega)$, we now compute the energy of the binary 
system
as a function of frequency. We define the binding energy $E_B$ of the binary
as,
\begin{align}
E_B\equiv H-M,
\end{align}
and use Eq.~\eqref{Hamiltonian} to evaluate the Hamiltonian. Using
Eq.~\eqref{eq:Kepler} to convert $r$ to $\Omega$ we plot the binding energy 
(normalized by the total mass $M$) as
a function of orbital frequency in Fig.~\ref{fig:Eomega}. We also plot the
binding energy computed from QE configurations in Ref.~\cite{Taniguchi:2014fqa}
as dashed lines and the 4PN prediction for nonspinning point
particles in GR~\cite{Damour:2014jta} as a green
dashed-dotted line.\footnote{We use the 4PN binding energy in GR as our 
benchmark rather than more sophisticated estimates for simplicity. For the 
frequency range we consider, the 4PN energy is visually indistinguishable from 
the predictions of the EOB formalism~\cite{Bohe:2016gbl}.} To improve 
comparison, we have added
to the predictions of our effective-action model (computed at Newtonian order),
the 1PN, 2PN, 3PN, and 4PN corrections to the binding energy in GR.  These 
corrections raise the binding energy closer to the other
curves in Fig.~\ref{fig:Eomega}, but do not influence the ordering of the various
curves, and thus do not affect our conclusions.

As expected, prior to the onset of dynamical scalarization, the binding energy
closely resembles that of the corresponding system in GR.
After dynamical scalarization occurs, we find significant differences between
our analytic model and the QE results of Ref.~\cite{Taniguchi:2014fqa}: the
present model predicts an increase in the magnitude of the binding energy
$|E_B|$ relative to GR whereas the QE computations indicate that the
magnitude should decrease.  Given the interpretation of dynamical scalarization
as a phase transition detailed in Sec.~\ref{sec:Landau}, one expects the
scalarized binary to be more tightly bound than the corresponding unscalarized
binary, i.e., the GR prediction. If this were not the case, dynamical
scalarization would be an endothermic process (requiring energy input) and the
$\varphi\rightarrow -\varphi$ symmetry would not spontaneously break. Based on
this intuition, the predictions of our model in Fig.~\ref{fig:Eomega} appear
qualitatively correct. The cause of the disagreement between our model and
Ref.~\cite{Taniguchi:2014fqa} remains unclear. The discrepancy could stem from
the assumption of conformal flatness and/or the presence of tidal interactions
absent in our point-particle model of the dynamics. However, to explain the
disagreement in Fig.~\ref{fig:Eomega}, these factors would need to play a more
significant role in the presence of scalar charges; analogous calculations done
in GR agree with analytic point-particle predictions
of the binding energy much more closely than the deviations shown in
Fig.~\ref{fig:Eomega}  (see, e.g., Ref.~\cite{Taniguchi:2010kj}).

\section{Dynamical scalarization as a phase transition}\label{sec:Landau}

Having validated its accuracy in Sec.~\ref{sec:results}, in this section we explore 
an important conceptual implication of our effective action model: we definitively 
establish dynamical scalarization as a second-order phase transition. Using the 
Landau theory of phase transitions~\cite{Landau:1937,*LandauTranslation}, we 
discuss the scalarization of an isolated NS (spontaneous scalarization), an 
equal-mass BNS (dynamical scalarization), and an unequal-mass BNS 
(spontaneous, induced, and dynamical scalarization).

The approach by Landau~\cite{Landau:1937,*LandauTranslation} allows one to 
relate certain types of phase transitions to broken symmetries. We begin with a 
schematic review, closely following Ref.~\cite{Landau:1937,*LandauTranslation}. 
Consider a system described by a set of state variables $\bm{\zeta}$ and 
thermodynamic potential $\Xi(\bm{\zeta})$ that undergoes a second-order 
transition between two phases at some critical point $\bm{\zeta}^*$. The degree 
of symmetry in each phase can be described by an order parameter $\eta$. We 
choose the order parameter such that it vanishes for the phase with greater 
symmetry, but in the other phase, the breaking of some of these symmetries 
causes $\eta$ to be nonzero. To exhibit a second-order phase transition, the 
thermodynamic potential must admit an expansion near the critical point of the 
form
\begin{align}
\Xi(\bm{\zeta},\eta)=\Xi_0(\bm{\zeta})+\Xi_2(\bm{\zeta}) \eta^2+\Xi_4(\bm{\zeta}) 
\eta^4+\Ord\left(\eta^6\right),\label{eq:XiExpand}
\end{align}
where the coefficients obey the following conditions:
\begin{align}
\Xi_4(\bm{\zeta})>&0,\label{eq:Cond1}\\
\Xi_2(\bm{\zeta}^*)=&0.\label{eq:Cond2}
\end{align}
The first condition guarantees that the system has an equilibrium solution (found 
at the minimum of $\Xi$). We discuss the second condition below.

For states ``above'' $\bm{\zeta}^*$, i.e., those for which $\Xi_2(\bm{\zeta})>0$, 
the potential~\eqref{eq:XiExpand} is positive definite, and so the system reaches 
equilibrium in the more symmetric state (the one in which $\eta$ vanishes). 
However, as one passes through the point $\bm{\zeta}^*$, the coefficient $
\Xi_2(\bm{\zeta})$ changes sign; now the potential~\eqref{eq:XiExpand} is 
minimized for configurations with nonzero values of $\eta$.

In anticipation of later discussion, we generalize the treatment above to systems described by a vector order parameter $\bm{\eta} \in \mathbb{R}^n$, where Euclidean coordinates are denoted with unitalicized Latin indices. In this generalization, the functions $\Xi_m(\bm{\zeta})$ become rank-$m$ tensors of dimension $n$ such that Eq.~\eqref{eq:XiExpand} becomes
\begin{align}
\begin{split}
\Xi(\bm{\zeta},\bm{\eta})=&\Xi_0(\bm{\zeta})+\left[\Xi_2(\bm{\zeta})\right]_\text{ab} \eta^\text{a} \eta^\text{b}+\left[\Xi_4(\bm{\zeta})\right]_\text{abcd} 
    \eta^\text{a}\eta^\text{b}\eta^\text{c}\eta^\text{d}\\
    &+ {\cal O}\left(\bm{\eta}^6\right) .\label{eq:XiExpandVector}
    \end{split}
\end{align}

The conditions~\eqref{eq:Cond1} and~\eqref{eq:Cond2} must be appropriately extended, as well. To ensure that the system has an equilibrium solution, we require that $\Xi_4$ be positive definite, in the sense that
\begin{align}
\left[\Xi_4(\bm{\zeta})\right]_\text{abcd} 
\eta^\text{a}\eta^\text{b}\eta^\text{c}\eta^\text{d} > 0, \qquad \forall \,  {\bm \eta} \in \mathbb{R}^n. \label{eq:Cond1Vector}
\end{align}
The $n$-dimensional generalization of Eq.~\eqref{eq:Cond2} is
\begin{align}
\det \left(\left[\Xi_2(\bm{\zeta}^*)\right]_\text{ab} \right)= 0.\label{eq:Cond2Vector}
\end{align}
Note that in the phase with greater symmetry, our assumption that $\Xi$ is minimized when $\bm{\eta}$ vanishes ensures that all eigenvalues of the matrix $\left[\Xi_2(\bm{\zeta}^*)\right]_\text{ab}$ must be positive. In the less symmetric phase, at least one of the eigenvalues must be negative; however, the determinant of the matrix remains positive if an even number of eigenvectors have negative eigenvalues.

\subsection{Spontaneous scalarization of an isolated body}

The classical illustration of a second-order phase transition is spontaneous 
magnetization in a ferromagnet at the Curie temperature $T_C$. In this example, 
$\Xi$ is the energy $E$ of the system and $\bm{\zeta}$ represents the 
temperature and external magnetic field $\bm{B}$. The order parameter $\eta$ is 
the total magnetization $\bm{M}\equiv  - \partial E / \partial \bm{B}$, which is 
thermodynamically conjugate to $\bm{B}$. Inspired by this example, 
\citet{Damour:1996ke} considered a phenomenological model of spontaneous 
scalarization following the Landau ansatz~\eqref{eq:XiExpand}. Starting from the 
total energy of an isolated NS $m_E(\varphi)$, the authors selected the potential 
$m(Q)$, defined as in Eq.~\eqref{Legendre}, to play the role of $\Xi$. The bulk 
properties of the NS are its baryonic mass $\bar m$ and external scalar field $
\varphi$. Analogous to spontaneous magnetization, the authors identified the 
order parameter $Q$ as the conjugate variable to the scalar field [c.f. 
Eq.~\eqref{eq:QmE}].\footnote{The notation of Ref.~\cite{Damour:1996ke} differs 
from that used here. The original notation can be recovered with the following 
substitutions:
${\varphi\rightarrow \varphi_0,}\  {Q\rightarrow \omega_A,}\ {m_E(\varphi)
\rightarrow m_A(\omega_A,\varphi_0),}\  {m(Q)\rightarrow \mu(\omega_A)}$.} 
The behavior of the potential $m$ around the critical baryonic mass $\bar{m}_
\text{cr}$ was modeled by~\cite{Damour:1996ke}
\begin{align}\label{eq:DEFLandauModel}
m(Q)=\frac{1}{2} a \left(\bar{m}_\text{cr}-\bar{m}\right) Q^2+\frac{1}{4} b Q^4,
\end{align}
where $a$ and $b$ are constant (positive) coefficients. Above the critical 
baryonic mass, NSs equilibrate in configurations with nonzero scalar charge.

By design, our
point-particle model~\eqref{mexpand2} takes the same form as
Eq.~\eqref{eq:DEFLandauModel}, and thus can model spontaneous
scalarization as well. Unlike Eq.~\eqref{eq:DEFLandauModel}, we do not factor 
out any mass-dependence of the coefficients $c^{(i)}$. As demonstrated in 
Section~\ref{sec:match}, our model remains valid for stars with $\bar{m} \not
\approx \bar{m}_\text{cr}$---these stars were not considered in Ref.~
\cite{Damour:1996ke}. The coefficient $c^{(2)}$ plays the role of $\Xi_2$ in
the Landau ansatz~\eqref{eq:XiExpand}; note that this coefficient depends on
the properties of the NS (e.g., the mass and EOS) and on the scalar-tensor 
coupling (characterized by $\beta$). The critical point at which a NS transitions 
from an unscalarized state ($Q=0$) to a spontaneously scalarized state ($Q
\neq0$) occurs when $c^{(2)}$ is zero. Neutron stars with negative values of 
$c^{(2)}$ must spontaneously scalarize; the unscalarized state is unstable.

\subsection{Dynamical scalarization of equal-mass binaries}\label{sec:DSPhaseTransition}

With our effective action model, we can now apply this analysis to a binary 
system of NSs. For simplicity, we begin by studying equal-mass systems with 
zero background scalar field $\varphi_0$. We assume that NSs have the same 
properties as well, i.e., $c^{(i)}_A = c^{(i)}_B$. For illustrative purposes, we drop 
the $\mathbf{p}^2$ and $m/r$ terms in the Hamiltonian~\eqref{Hamiltonian}; 
restoring these terms does not affect the qualitative behavior we describe below.

Under these assumptions, the Hamiltonian is given by
\begin{align}\label{eq:H:equalmass}
H = 2 m^{(0)} + \left(c^{(2)}-\frac{1}{r}\right) Q^2 + \frac{c^{(4)}}{12} Q^4,
\end{align}
where we have dropped the body labels. This expression takes the same form as 
Eq.~\eqref{eq:XiExpand}. Using the same analysis as in the previous subsection, 
we show that dynamical scalarization is a second-order phase transition that 
occurs at a separation ${r_\text{DS}=1/c^{(2)}}$; this conclusion agrees with our 
prediction in Eq.~\eqref{eq:Q:solution:equalmass}. By comparing Eqs.~\eqref{mexpand2} 
and~\eqref{eq:H:equalmass} we see that an equal-mass 
dynamically scalarizing system behaves like an isolated NS with an effective 
coefficient $c^{(2)}_\text{eff}\equiv c^{(2)}-1/r$ that decreases as the binary 
coalesces.

\begin{figure}
\includegraphics[width=\columnwidth,clip=true, trim= 15 0 15 25]{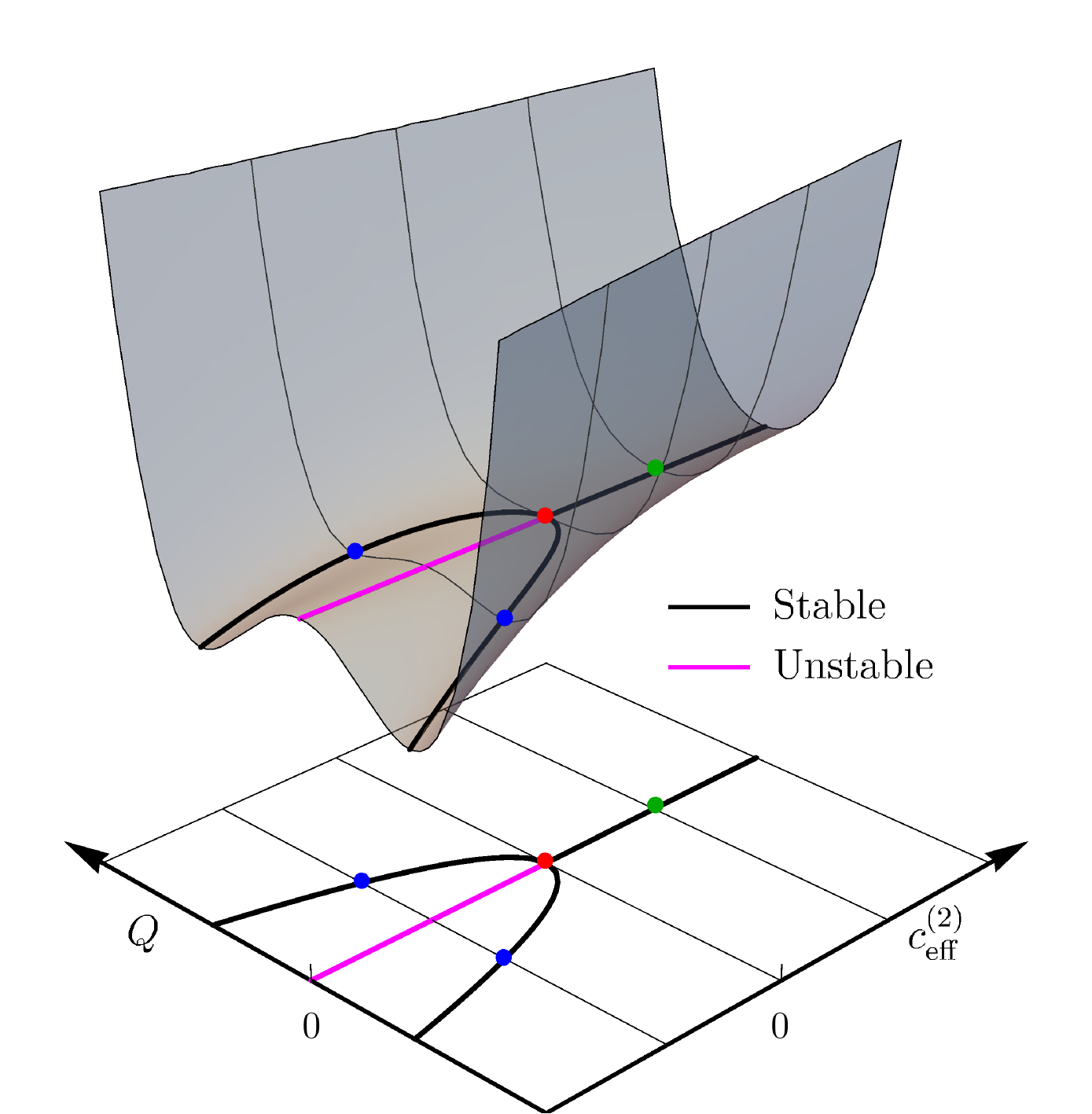}
\caption{Illustration of the Hamiltonian of an equal-mass BNS as function of 
scalar charge and effective
  coefficient $c^{(2)}_\text{eff}=c^{(2)}-1/r$. Solutions to the equations of
  motion are highlighted with solid lines. When $c^{(2)}_\text{eff}$ becomes
negative, the trivial solutions $Q=0$ become unstable. The bottom lower plane
shows the projection of the solutions.} \label{fig:SurfPlot}
\end{figure}

In Fig.~\ref{fig:SurfPlot}, we plot the simplified Hamiltonian~
\eqref{eq:H:equalmass} as a function of charge and effective coefficient 
$c^{(2)}-1/r$. For positive values of this effective coefficient, the energy is 
minimized in the trivial configuration $Q=0$. Below the critical point $c^{(2)}_
\text{eff}=0$, the unscalarized state becomes unstable; instead, the binary 
system transitions into a scalarized state. The bottom plane shows the projection 
of the equilibrium solutions in black. As predicted by Eq.~
\eqref{eq:Q:solution:equalmass}, the stable solutions bifurcate at the critical 
point, spontaneously breaking the scalar-parity symmetry of the theory. Note that 
this entire discussion can be applied directly to isolated NSs that undergo 
spontaneous scalarization by taking $r\rightarrow \infty$.

\subsection{Scalarization of unequal-mass binaries}

\begin{figure*}
\includegraphics[width=\textwidth, clip=true, trim= 0 15 55 10]{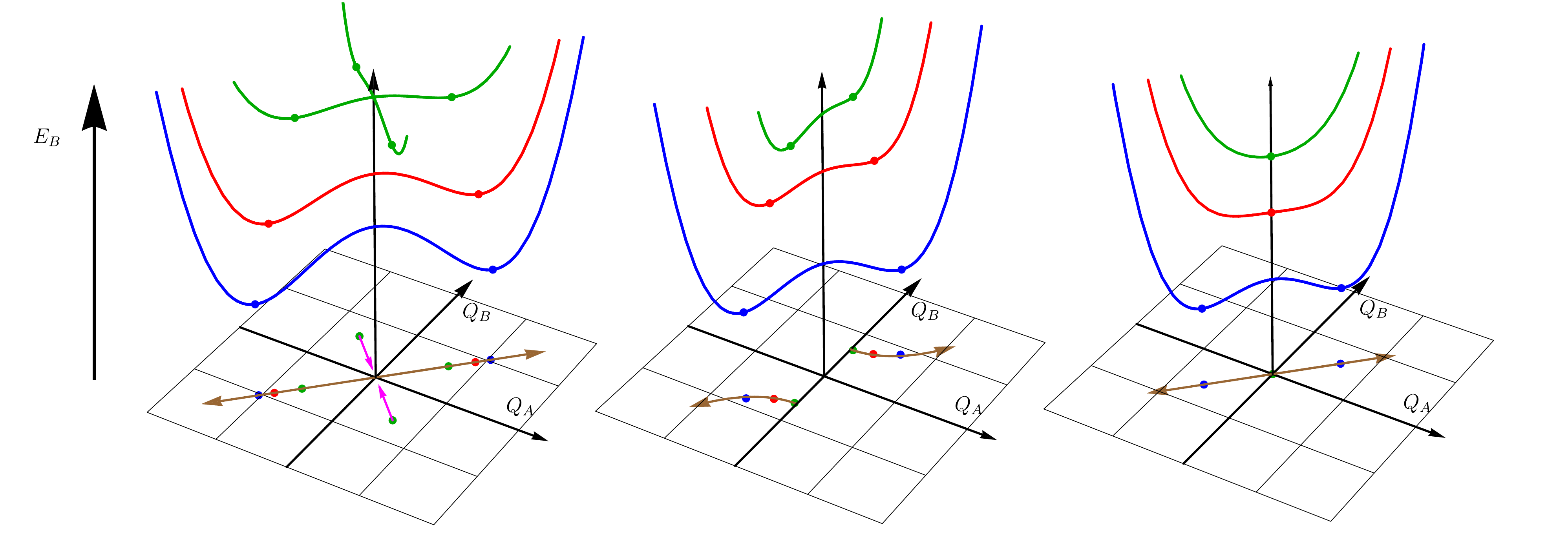}
\caption{Illustration of the binding energy as a function of scalar charge for BNSs 
that undergo: (left) spontaneous scalarization, (middle) induced scalarization, 
and (right) dynamical scalarization. Equilibrium solutions are highlighted with 
dots. The solutions are projected onto the $(Q_A,Q_B)$ plane below; colored 
arrows indicate the flow of these solutions as the binary coalesces.} 
\label{fig:ThreeCurvePlot}
\end{figure*}

Finally, we turn our attention to the critical phenomena that can occur in 
unequal-mass binaries. The (vector) order parameter $\bm{\eta}\in \mathbb{R}^2$ is given by
\begin{align}
(\eta^1,\eta^2)= (Q_A,Q_B) \,.
\end{align}
Again, we assume that the background scalar field $\varphi_0$ vanishes and drop the $\mathbf{p}^2$ and $m/r$ terms in the Hamiltonian~\eqref{Hamiltonian}; these simplifications do not affect the qualitative behavior described below. Under these assumptions, the Hamiltonian takes the same form as Eq.~\eqref{eq:XiExpandVector} with
\begin{align}
\Xi_0=&m_A^{(0)}+m_B^{(0)},\label{eq:Xi0:UnequalMass}\\
\left[\Xi_2\right]_\text{ab}=& \frac{1}{2}\begin{pmatrix} c_A^{(2)}& -r^{-1}\\ -r^{-1} & c_B^{(2)} \end{pmatrix},\label{eq:Xi2:UnequalMass}\\
\left[\Xi_4\right]_\text{abcd}=&\frac{1}{24}\left(c_A^{(4)}\delta_\text{a}^{1}\delta_\text{b}^{1}\delta_\text{c}^{1}\delta_\text{d}^{1}+c_B^{(4)}\delta_\text{a}^{2}\delta_\text{b}^{2}\delta_\text{c}^{2}\delta_\text{d}^{2}\right).\label{eq:Xi4:UnequalMass}
\end{align}

We examine the Hamiltonian~\eqref{Hamiltonian} for 
systems that undergo:
\begin{enumerate}
\item \textit{Spontaneous scalarization}: Both stars are initially scalarized $
(c^{(2)}_A<0,\,c^{(2)}_B<0)$,
\item \textit{Induced scalarization}: Only one star is initially scalarized $(c^{(2)}
_A>0,\,c^{(2)}_B<0)$,
\item \textit{Dynamical scalarization}: Neither star is initially scalarized $(c^{(2)}
_A>0,\,c^{(2)}_B>0)$.
\end{enumerate}
For all three cases, we restrict our attention to binaries following circular orbits. 
The binding energy is shown in Fig.~\ref{fig:ThreeCurvePlot} as a function of the 
NS charges. We show only the slices of the full graph $E_B(Q_A,Q_B)$ that pass 
through equilibrium solutions; for comparison, these curves correspond to the 
thin black lines on the surface in Fig.~\ref{fig:SurfPlot}. Moving from left to right, 
the plots correspond to spontaneous, induced, and dynamical scalarization, 
respectively. Moving downwards in each plot, the green, red, and blue curves 
depict the binding energy at progressively smaller separations. The equilibrium 
solutions are denoted with dots on the curves and are projected onto the $
(Q_A,Q_B)$-plane in the color corresponding to their separation. The colored 
arrows depict the flow of equilibrium solutions as the separation decreases.

At large separations (green), there exist four stable configurations for 
spontaneously scalarized binaries (left panel): each NS can exhibit a positive or 
negative scalar charge, and the choices for each are uncorrelated. However, as 
the separation decreases (red and blue), configurations in which the two stars 
have opposite-parity charges become energetically unfavorable. As indicated by 
the pink arrows, these solutions flow towards the origin and transform into a 
saddle point, i.e., this branch of solutions becomes unstable. Thus, at this critical 
separation (red) there exists a new phase transition distinct from those discussed 
above. From Eqs.~\eqref{eq:Cond2Vector} and~\eqref{eq:Xi2:UnequalMass}, we find that this critical point occurs at a separation of $r^*=(c_A^{(2)} c_B^{(2)})^{-1/2}$. Unlike with dynamical scalarization, the more symmetric state phase occurs at separations smaller than $r^*$. The equilibrium solutions with charges of the same sign flow away from 
the origin as the binary coalesces. The charge of each spontaneously scalarized 
star will continue to grow during the inspiral due to feedback from its companion.

Binaries that undergo induced scalarization (middle panel) begin with an 
unscalarized star $Q_A=0$ and a scalarized star $Q_B\neq0$ (green). As the 
stars are brought closer together (red), the unscalarized star rapidly develops 
scalar charge, whereas the initially scalarized star remains (approximately) 
unchanged. However, as the separation decreases further (blue), the two 
charges become of the same order of magnitude and continue to increase at 
roughly the same rate through the remainder of the coalescence. Unlike for 
spontaneous and dynamical scalarization, the branches of equilibrium solutions 
are disjoint throughout the entire coalescence, i.e. the colored arrows in 
Fig.~\ref{fig:ThreeCurvePlot} never meet. Because $c_A^{(2)}$ and $c_B^{(2)}$ have opposite signs, the determinant of $\left[\Xi_2\right]_\text{ab}$ [given in Eq.~\eqref{eq:Xi2:UnequalMass}] is negative for all separations. Induced scalarization fails to meet condition~\eqref{eq:Cond2Vector} and therefore cannot be classified as a phase transition.

Finally, initially unscalarized unequal-mass binaries (right panel) evolve similarly 
as in Fig.~\ref{fig:SurfPlot}. As seen in Fig.~\ref{fig:ThreeCurvePlot}, the binary 
system begins in an unscalarized state (green). At the critical transition point 
(red), the effective $c^{(2)}$ coefficient vanishes; beyond that point (blue), 
scalarization becomes energetically favorable. Again, Eqs.~\eqref{eq:Cond2Vector} and~\eqref{eq:Xi2:UnequalMass} reveal that dynamical scalarization occurs at a separation of $r_\text{DS}=(c_A^{(2)} c_B^{(2)})^{-1/2}$, which reduces to the result in Sec.~\ref{sec:DSPhaseTransition} when $c_A^{(2)}= c_B^{(2)}$. As before, the scalar charges 
continue to grow after the onset of dynamical scalarization. 

\section{Conclusions}\label{sec:Conclusions}

In the present paper, we developed a new point-particle model for NSs in 
scalar-tensor gravity that can reproduce spontaneous, induced, and dynamical 
scalarization. The model parametrizes the various scalarization phenomena by 
just two coefficients $c^{(2)}$, $c^{(4)}$ for each NS. This approach should be 
contrasted with previous analytic models of dynamical scalarization~
\cite{Palenzuela:2013hsa,Sennett:2016rwa}, which relied upon numerically 
solving equations containing the generic function $m_E(\varphi)$. For the first 
time, we have computed a two-body Hamiltonian that incorporates dynamical 
scalarization in a self-consistent manner (see Ref.~\cite{Sennett:2016rwa} for a 
discussion of previous attempts). Observables derived from the model at
leading order in the PN expansion were shown to be in
good agreement with earlier analytic models and numerical QE calculations.
The identification of the relevant dynamical variables
in the effective action is crucial to our model.

Analogous to the analysis done in Ref.~\cite{Damour:1996ke} concerning 
spontaneous scalarization, our model rigorously establishes dynamical 
scalarization as a phase transition as per Landau theory~\cite{Landau:1937,
*LandauTranslation}. Additionally, it demonstrates the intimate connection 
between spontaneous and dynamical scalarization. The mapping between an 
equal-mass BNS undergoing dynamical scalarization and an effective 
spontaneously scalarized NS is detailed in Sec.~\ref{sec:DSPhaseTransition}.

Our effective action stands as an important first step towards accurate analytic 
waveforms of dynamically scalarizing BNSs. The model benefits from its close 
analogy to the effective action model of dynamical tides detailed in 
Refs.~\cite{Hinderer:2016eia, Steinhoff:2016rfi}---the dynamical scalar 
monopole $Q$ here corresponds to the dynamical gravitational quadrupole
 therein. References~\cite{Hinderer:2016eia, Steinhoff:2016rfi} derived an accurate
EOB~\cite{Buonanno:1998gg, Buonanno:2000ef} waveform model incorporating 
dynamical tidal
interactions. Using this model as a template, one could construct an analogous 
model for dynamical scalar-tensor effects. This construction will require 
calculations of dissipative effects and higher PN order results for the 
conservative dynamics.

Another avenue for future work is the addition of 
kinetic-energy terms to the effective action as in Eq.~\eqref{Smosci}. Resonant 
effects play an important role in the dynamical tides model of 
Refs.~\cite{Hinderer:2016eia, Steinhoff:2016rfi}; it remains to be seen whether analogous
effects could be important with dynamical scalar charges. Formulating the effective 
action in this manner offers a conceptual advantage over the current model, as it 
guarantees that all of the equations of motion are ordinary differential equations 
(rather than a mix of nonlinear algebraic and differential equations).

Finally, an intriguing extension of this work is to theories with a massive scalar field. Pulsar timing cannot constrain sufficiently short-range scalar fields, so a much wider range of parameter space of massive scalar-tensor theories remains to be constrained by GW observations than that of theories with a massless scalar~\cite{Ramazanoglu:2016kul}. The PN dynamics of a simple massive scalar-tensor theory were investigated in Ref.~\cite{Alsing:2011er}, and spontaneous scalarization of isolated NSs was studied in Refs.~\cite{Ramazanoglu:2016kul, Yazadjiev:2016pcb}; the framework we have presented above could synthesize these results with appropriate modifications to the field equations~\eqref{eq:EinsteinEq} and~\eqref{eq:KleinGordonEq}.

\acknowledgments

We are grateful to Andrea Taracchini for discussions and Alessandra Buonanno
for helpful comments.

\bibliography{refs.bib}

\end{document}